\documentclass[12pt, prd, showpacs]{revtex4}
\usepackage{amssymb}
\usepackage{amsmath}

\input{tcilatex}

\begin{document}

\title{Acceleration of particles by quasiblack holes}
\author{O. B. Zaslavskii}
\affiliation{Kharkov V.N. Karazin National University, 4 Svoboda Square, Kharkov, 61077,
Ukraine}
\email{zaslav@ukr.net}

\begin{abstract}
If two particles collide near the black hole horizon, the energy in the
centre of mass (CM) frame can grow indefinitely (the so-called the BSW
effect). This requires fine-tuning the parameters (the energy, angular
momentum or electric charge)\ of one particle. We show that \ the CM energy
can be unbound also for collisions in the space-time of quasiblack holes -
QBHs (the objects on the threshold of forming the horizon which do not
collapse). It does not require special fine-tuning of parameters and occurs
when any particle inside a QBH having a finite energy collides with the
particle that entered a QBH from the outside region.
\end{abstract}

\maketitle



\section{Introduction}

If two particles collide near the black hole horizon, the energy in their
centre of mass (CM) frame can grow indefinitely. This interesting effect was
discovered by Ba\~{n}ados, SIlk and West \cite{ban} and is called the BSW
effect after the names of its authors and is under active study now. To
achieve this effect, the collision by itself is insufficient. Additionally,
one of particles has to have fine-tuned parameters, so that the some
relationship between its energy and angular momentum or charge should hold.
This somewhat restricts the realization of the BSW effect or requires
special scenarios like collisions near circular orbits (that can occur in
the accretion disc) \cite{circ}. The aim of the present work is point out
that there exists one more class of objects - so-called quasblack holes
(QBH) - for which the BSW effect can reveal itself. What is especially
interesting is that for QBHs no fine-tuning is required at all, so the BSW
effect is even more universal in this sense than one could think before.

QBH is an object, roughly speaking, on the verge of collapsing and forming
the event horizon but without actual forming the horizon. This includes a
quite diverse set of physical objects: the self-gravitating monopole,
so-called extremal dust and even a usual thin shell in the near-horizon
limit (see \cite{qbh} and references therein for more strict definition and
discussion of general properties). Especially interesting is the fact that
QBHs realize the limiting transition from perfectly rotating fluid \cite%
{mein1} or a rapidly rotating disc \cite{disc} to the extremal black hole
that can have important consequences in relativistic astrophysics. However,
as the rotating (more realistic) case is more complicated, it looks
reasonable to analyze the spherically symmetric space-time. It is expected
to posses qualitatively similar features since rotation and charge have much
in common in black hole physics.\ Meanwhile, it is much more easy for
analysis. We do not touch upon here the issue of stability of QBHs which is
a separate question. We only point out that at present, there are some
preliminary results that lend support to the idea that due to the pressure,
QBHs can be stable and correspond to frozen stars \cite{pres}.

There is also additional motivation for studying the BSW effect in this
context. There are works in which the analogue of the BSW effect is
considered when the horizon is absent \cite{pj}, \cite{p}. The corresponding
approach is essentially based on effects of self-gravitation of moving
shells, so it looks more simple to reveal some features of the BSW effect
without horizons in a more simple situation.

In the present paper I restrict myself with the radial motion of charged
particles in spherically symmetric metric that represents the counterpart of
similar kind of the BSW effect for rotating black holes \cite{jl}.\
Meanwhile, the rotational analogue of QBHs also exists \cite{rot}, so the
results can be extended to it in a straightforward manner.

\section{BSW effect for black holes}

Let us remind the main ingredients of the BSW effect for charged spherically
symmetric black holes \cite{jl}. The metric can be written in the form%
\begin{equation}
ds^{2}=-f(r)dt^{2}+\frac{dr^{2}}{A(r)}+r^{2}d\omega ^{2}\text{.}  \label{m}
\end{equation}%
In the particular case of the Reissner-Nordstr\"{o}n (RN) metric $f=A=1-%
\frac{2M}{r}+\frac{Q^{2}}{r^{2}}$ where $M$ is the mass of the black hole, $%
Q $ is its charge, the electric potential $\varphi =\frac{Q}{r}$. We use the
systems of units in which the fundamental constants $G=c=$%
h{\hskip-.2em}\llap{\protect\rule[1.1ex]{.325em}{.1ex}}{\hskip.2em}%
=1. I consider, for definiteness, the extremal case, so for the RN black
hole $M=Q=r_{+}$ where $r_{+}$ is the horizon radius, $f=\left( 1-\frac{r_{+}%
}{r}\right) ^{2}$. The equations of motion give us 
\begin{equation}
mu^{0}=m\dot{t}=\frac{1}{f}(E-q\varphi )\text{,}  \label{u0}
\end{equation}

\begin{equation}
m^{2}\dot{r}^{2}=\frac{A}{f}(E-q\varphi )^{2}-m^{2}A\text{,}  \label{ur}
\end{equation}%
where dot denotes derivative with respect to the proper time of a moving
particle. Here, $m$ is the mass of a particle, $q$ is its electric charge, $%
E $ is its conserved energy.

Then, one can define the energy in the CM frame $E_{c.m.}$ according to $%
E_{c.m.}^{2}=-P^{\mu }P_{\mu }$ where $P^{\mu }=p_{1}^{\mu }+p_{2}^{\mu }$
is the total momentum, $p^{\mu }$ are individual ones of particles,
subscript $i$ enumerates particles' characteristics ($i=1,2$)$.$ Assuming
for simplicity that $m_{1}=m_{2}=m$, one finds that%
\begin{equation}
\frac{E_{cm}^{2}}{2m^{2}}=1+\frac{X_{1}X_{2}-Z_{1}Z_{2}}{N^{2}m^{2}}
\label{exz}
\end{equation}%
where 
\begin{equation}
X_{i}=E_{i}-q_{i}\varphi \text{, }Z_{i}=\sqrt{X_{i}^{2}-m^{2}N^{2}}
\label{xz}
\end{equation}%
and we introduce the lapse function, $f=N^{2}\,.$

Let particle 1 be critical and particle be usual. By definition, this means
that $\left( X_{1}\right) _{+}=0$, hereafter subscript "+" denotes
quantities calculated on the horizon. Then, 
\begin{equation}
E_{1}=q_{1}\varphi (r_{+})  \label{cr}
\end{equation}%
and $\left( X_{2}\right) _{+}\neq 0$. If $\left( X_{1}\right) _{+}$ does not
vanish exactly but is small in some sense (see below), we will call such a
particle "near-critical".

Near the horizon, we can use the Taylor expansion for the potential $\varphi
\approx \varphi (r_{+})[1-a\frac{(r-r_{+})}{r_{+}}]+O((r-r_{+})^{2})$ and
the lapse function $N\approx b\frac{(r-r_{+})}{r_{+}}$ where we took into
account that for the extremal horizon, by definition, $f\sim (r-r_{+})^{2}$.
In the RN case, the constants $a=b=1$.

Then, if is easy to obtain from (\ref{xz}), (\ref{cr}) that for a critical
particle 
\begin{equation}
X_{1}\approx \frac{a}{b}E_{1}N\text{, }
\end{equation}%
near the horizon.

Then, in the limit $f\rightarrow 0$ one can find that%
\begin{equation}
E_{cm}^{2}\approx \frac{2X_{2(H)}}{N}[\frac{a}{b}E_{1}-\sqrt{(\frac{a}{b}%
)^{2}E_{1}^{2}-m^{2}}]\text{.}
\end{equation}

Thus the CM energy grows unbound that gives us the BSW effect.

\section{Case of quasiblack holes}

In recent years, a new physical object was singled out as a separate
physical entity - so-called quasiblack holes. Roughly speaking, a quasiblack
hole is a system on the threshold of formation of the event horizon when a
system can approach this threshold as close as one likes but actually the
horizon does not form. This includes a number of physical objects -
self-gravitating Higgs magnetic monopoles \cite{lw1}, \cite{lw2}, extremal
charged dust \cite{bon}, \cite{je} and even simple extremal shells \cite{qbh}%
. More rigorous definition, brief review of existing realizations with
extended list of references and discussion of general properties of
quasblack holes can be found in \cite{qbh}, see also recent work \cite{rec}.

To formulate the problem in mathematical terms, let us consider a compact
body for which $N(r_{B})=\varepsilon $ is small on the boundary $r=r_{B}$.
We assume that inside ($r<r_{B}$)%
\begin{equation}
N=\varepsilon \tilde{N}(r)
\end{equation}%
where $\tilde{N}(r)\neq 0$, $\tilde{N}(r_{B})=1$ because of continuity of
the metric on the boundary. Eq. (\ref{exz}) is still valid. Inside, one can
introduce a new time coordinate according to $\tilde{t}=\varepsilon t$.
Then, in the inner region the metric takes the form 
\begin{equation}
ds^{2}=-\tilde{N}^{2}(r)d\tilde{t}^{2}+\frac{dr^{2}}{A(r)}+r^{2}d\omega ^{2}%
\text{.}
\end{equation}%
The small parameter $\varepsilon $ is eliminated, so $\tilde{t}$ is a
natural measure of time in the inner region.

Let particle 2 enter the inner region (say, from infinity) and collide there
with particle 1. We assume that particle 2 was created \textit{inside} the
boundary and has some finite energy $\tilde{E}$ measured with respect to the 
\textit{inner time }$\tilde{t}$. As the energy is a time component of the
four-vector, $E=\tilde{E}\frac{\partial \tilde{t}}{\partial t}=\varepsilon 
\tilde{E}$. In a similar way, $\varphi =\varepsilon \tilde{\varphi}$. As a
result, $X_{1}=\varepsilon \tilde{X}_{1}$ where $\tilde{X}_{1}$ does not
contain a small parameter.

It is clear from the formulas of the previous Section that it is smallness
of $\left( X_{1}\right) _{+}$ that causes indefinite growth of $E_{c.m.}$.
But this is just what happens now! Indeed, it follows from (\ref{exz}) that%
\begin{equation}
E_{cm}^{2}\approx \frac{2X_{2(H)}}{\varepsilon \tilde{N}^{2}}[\tilde{X}_{1}-%
\sqrt{\tilde{X}_{1}^{2}-m^{2}\tilde{N}^{2}}]\text{.}  \label{qe}
\end{equation}

Therefore, if $\varepsilon \ll 1,$ $E_{cm}^{2}$ can be made as large as one
likes.

\section{Discussion: comparison of two kinds of the BSW effect}

The existence of the BSW effects for black holes is due to the horizon: the
metric function $f=0$ at some point $r=r_{+}$. By contrast, in the QBH case,
the metric function does not vanish at all. Instead, it remains small but
nonzero in the entire region $r\leq r_{B}$. Correspondingly, the BSW effect
can occur there not only near the would-be horizon (like near the horizon of
black holes) but \textit{everywhere} inside. Actually, this is nontrivial
consequence of mismatch between the time coordinates inside and outside that
is important property of QBH \cite{qbh}.

We saw in the preceding Section that $X_{1}\rightarrow 0$ for any typical
particle created inside. In this sense, any such particle is near-critical.
Meanwhile, particle 2 is usual, so the main condition of the BSW effect is
satisfied. It is worth stressing that this condition is obtained without
fine-tuning at all, so \textit{any} particle inside with a finite energy $%
\tilde{E}$ is near-critical. As particle 2 is chosen to be usual, their
collision leads to the BSW effect.

It is also instructive to describe the situation in a kinematic language.
For black holes, a critical particle approaches the horizon with some finite
velocity $V_{1}<1$ measured by a static or stationary observer. Any usual
particle has $V_{2}\rightarrow 1$ near the horizon. As a result, the
relative velocity of two particles $w\rightarrow 1$, the corresponding
Lorentz gamma factor indefinitely grows and $E_{c.m.}\rightarrow \infty $
(see \cite{k} for details).

For radial motion in the metric (\ref{m}), $V=\frac{dl}{d\tau _{st}}$ where $%
dl$ is the proper distance and $d\tau _{st}=\sqrt{f}dt$ is the proper time
measured by a static observer in a given point. Using (\ref{m}), one finds 
\begin{equation}
V=\frac{\left\vert \dot{r}\right\vert }{\sqrt{\dot{r}^{2}+A}}\text{,}
\label{v}
\end{equation}%
In the inner region, $\tilde{N}$ $\ $and $\tilde{A}$ should be used here
instead of $N$ and $A$.

In the black hole case (say, for the RN metric), $A\sim f\rightarrow 0$ near
the horizon. If particle is usual, $\dot{r}^{2}\neq 0$ due to the first term
in the right hand side of (\ref{ur}), so $V\rightarrow 1$ for a usual
particle. In a similar way, for the critical one, $\dot{r}^{2}\sim
N^{2}\rightarrow 0$ and we have from (\ref{v}) that $V_{1}<1$.

For QBHs the situation is somewhat different. The quantity $A$ is small near
the boundary from the outside but is finite nonzero in the inner region
(say, we can take Minkowski \ vacuum inside and the RN metric outside) \cite%
{qbh}. The quantity $N\sim \varepsilon $ is small inside. Let us consider
the process from the viewpoint of an observer who resides inside the
boundary $r_{B}$ and uses time $\tilde{t}$. Particle 1 created just inside
with the finite energy $\tilde{E}$, has also finite $\dot{r}$, $\tilde{N}$, $%
\tilde{A}$. As a result, $V_{1}<1$. From the other hand, for a generic
particle 2 it follows from eq. (\ref{ur}) that $\left\vert \dot{r}%
\right\vert \rightarrow \infty $ since $A$ is finite but $f\sim \varepsilon
^{2}$ is small in the inner region. Therefore, it follows from (\ref{v})
that $V_{2}\rightarrow 1$. Thus, the mechanisms that give rise to $V_{1}<1$, 
$V_{2}\rightarrow 1$ in both cases are different but the result is the same:
the BSW effect due to the relative velocity approaching the speed of light.

Thus, we obtained by different methods that particle 1 in this context is
always near-critical, so due to the collision between a near-critical and
usual particles the BSW effect occurs.

The situation for which the BSW effect occurs can be summarized in the table.

\begin{tabular}{|l|l|l|l|l|l|l|}
\hline
Particle & Origination & $E$ & $\tilde{E}$ & $V$ inside & $\left\vert \dot{r}%
\right\vert $ inside & Type \\ \hline
1 & created inside & $\sim \varepsilon $ & finite & $<1$ & finite & 
near-critical \\ \hline
2 & coming from outside & finite & $\sim 1/\varepsilon $ & $\approx 1$ & $%
\sim 1/\varepsilon $ & usual \\ \hline
\end{tabular}

It is worth mentioning that a quite different mechanism for the BSW effect
for regular space-times without horizons was considered in \cite{pj}. It
applies to pure radial motion of uncharged particles, required some special
limitations on the velocities and angular momentums and leads to the
dependence $E_{cm}^{2}\sim f^{-1}=N^{-2}$ whereas in our case $%
E_{cm}^{2}\sim N^{-1}$ and the only physical requirement is the smallness of
the parameter $\varepsilon $.

Two reservations are in order. First, in the present paper, we neglected the
size of particles and treated them as point-like objects. If one takes into
account the size of falling objects, one should also take into account the
tidal forces acting on them. For $\varepsilon \rightarrow 0$, the tidal
forces for an observer falling inside the QBH grow unbound (see Sec. III 1
of \cite{qbh}). This prevents an object of a finite size to penetrate from
the outside to inside. But for sufficiently small size and small but nonzero 
$\varepsilon $ one can neglect this effect. More precisely, one can obtain
restriction similar to that obtained in \cite{cl} in another context.
Indeed, it was shown in \cite{qbh} that typical value of the Riemann tensor
in the free-falling frame has the order $R\sim \varepsilon ^{-2}M^{-2}$
where $M$ is a black hole mass. We require that, say, $R<\frac{1}{l_{0}^{2}}$%
, where $l_{0}$ is a typical scale of an object which can withstand
gravitational tidal forces. Then, taking for a typical black hole mass $%
M\sim $ 10$^{5}cm$ in geometrical units and $l_{0}=10^{-8}$ cm is an atomic
size one obtains that $\varepsilon >10^{-13}$ that is not restrictive at all.

Second, if an object of a finite size moves inside the QBH with matter its
trajectory is not geodesic. However, one can expect that this does not
affect our results significantly even if corrections due to nonzero size are
taken into account since (i) the energy of an incoming particle is much
larger than that inside since it contains a small parameter $\varepsilon $
(see the Table), (ii) it was shown in \cite{gc} that the BSW effect retains
its validity even for nongeodesic motion. It is worth also reminding that
one of realization of a QBH is a simple vacuum shell in which case there is
no matter inside at all.

\section{Summary and conclusions}

Thus we extended the class of objects which can be sources of the BSW
effect.\ In addition to extremal \cite{ban} or nonextremal \cite{gp} black
holes and naked singularities \cite{nk}, quasiblack holes are added and the
mechanism of creating the BSW effect is analyzed. It turned out that the BSW
effect may occur not only near the distinct surfaces but in the entire
region. In so doing, there is no need to achieve fine-tuning by special
selection of the particles for obtaining critical ones which are necessary
for the BSW effect. Therefore, the results obtained look quite universal and
can be useful in the analysis of physics of relativistic objects which are
close to the stage of gravitational collapse. The possibility to accelerate
particles is one more interesting property to QBHs in addition to those
discussed in \cite{qbh}. In the present paper we restricted ourselves by
radial motion in the background of spherically symmetric QBHs. Meanwhile,
the rotating analogue of QBH appears naturally in astrophysics of rapidly
rotating objects (see \cite{mein1} and references therein). Correspondingly,
the BSW effect should occur also for rotating QBHs.

\end{document}